\documentclass{aastex62}

\newcommand{\BaII}{\ion{Ba}{2}}
\newcommand{\Cratio}{$\rm^{12}C/^{13}C$}
\newcommand{\Hipp}{{\it Hipparcos}}

\newcommand{\Gaia}{{\it Gaia}}
\newcommand{\Stromgren}{Str\"omgren}
\newcommand{\kms}{\ifmmode~{\rm km~s}^{-1}\else ~km~s$^{-1}~$\fi}

\received{xxx}
\revised{xxx}
\accepted{xxx}
\submitjournal{ApJ}

\shorttitle{Carbon-Deficient Red Giants}
\shortauthors{Bond}

\begin{document}          

\title{Carbon-Deficient Red Giants}

\author{Howard E. Bond}  

\affil{Department of Astronomy \& Astrophysics, Pennsylvania State
University, University Park, PA 16802}
\affil{Space Telescope Science Institute, 3700 San Martin Drive,
Baltimore, MD 21218}
\affil{Visiting astronomer, Cerro Tololo Inter-American Observatory and Kitt Peak National Observatory, National Optical Astronomy Observatory, which are operated by the Association of Universities for Research in Astronomy under a cooperative agreement with the National Science Foundation.}

\begin{abstract}

Carbon-deficient red giants (CDRGs) are a rare class of peculiar red giants,
also called ``weak G-band'' or ``weak-CH'' stars. Their atmospheric compositions
show depleted carbon, a low \Cratio\ isotopic ratio, and an overabundance of
nitrogen, indicating that the material at the surface has undergone CN-cycle
hydrogen-burning. I present \Stromgren\ {\it uvby\/} photometry of nearly all
known CDRGs. Barium stars, having an enhanced carbon abundance, exhibit the
``Bond-Neff effect''---a broad depression in their energy distributions at
$\sim$4000~\AA, recently confirmed to be due to the CH molecule. This gives
\BaII\ stars unusually low \Stromgren\ $c_1$ photometric indices. I show that
CDRGs, lacking CH absorption, exhibit an ``anti-Bond-Neff effect'': higher 
$c_1$ indices than normal red giants. Using precise parallaxes from \Gaia\/ DR2,
I plot CDRGs in the color-magnitude diagram (CMD) and compare them with theoretical
evolution tracks. Most CDRGs lie in a fairly tight clump in the CMD, indicating
initial masses in the range $\sim$2 to $3.5\,M_\odot$, if they have evolved as
single stars. It is unclear whether they are stars that have just reached the
base of the red-giant branch and the first dredge-up of CN-processed material,
or are more highly evolved helium-burning stars in the red-giant clump. About
10\% of CDRGs have higher masses of $\sim$4 to $4.5\,M_\odot$, and exhibit unusually
high rotational velocities. I show that CDRGs lie at systematically larger
distances from the Galactic plane than normal giants, possibly indicating a role
of binary mass-transfer and mergers. CDRGs continue to present a major puzzle
for our understanding of stellar evolution.

\end{abstract}

\keywords{stars: abundances --- stars: atmospheres --- stars: chemically
peculiar --- stars: evolution}

\section{Red Giants with Abnormally Weak CH Absorption}

The unusual spectrum of the fifth-magnitude star HR~885 was noted on a Harvard
objective-prism plate by Annie~J. Cannon (1912). She described it as having a
late-type spectrum that was peculiar in showing a weakened G~band of the CH
molecule near 4300~\AA: unlike the normally continuous absorption at the G~band,
the feature was separated into several individual absorption lines. HR~885 was
included in the Mt.~Wilson catalog of spectroscopic absolute magnitudes (Adams
et al.\ 1935), and assigned a spectral type of G4 and a visual absolute
magnitude of +0.2, without further comment. However, Bidelman (1951), based on a
slit spectrogram, remarked that ``the spectrum is extraordinarily peculiar, with
the line spectrum matching fairly well G5~III but with no trace of CN or CH
absorption.'' Bidelman suggested that HR~885 may present a ``unique case of {\it
low\/} carbon abundance.''

HR 885 is the prototype of a class of peculiar stars that have been called
``weak-CH'' or ``weak-G~band'' stars. Given that all known members of this class
have proven to be red giants, in this paper I will use a more astrophysical
terminology of ``carbon-deficient red giants'' (CDRGs).

Another star in which ``CH is very weak,'' 37~Com, was noted by Roman (1952) in
her spectral reconnaissance of bright northern F5--K5 stars. An initial
curve-of-growth abundance analysis of HR~885, and a similar but less extreme
CDRG, HR~6791,\footnote{An extensive and useful historical discussion of this
star, other CDRGs, and their peculiarities is given by Griffin (1992).} was
carried out by Greenstein \& Keenan (1958). Both stars were confirmed to be
underabundant in carbon, by an average factor of about 20, but with
approximately normal metal content. 

% In HD~885, CN and CH are deficient by factors of $\sim\!30$ and $\sim\!100$,
% respectively.

% Griffin refers to early abund studies:
% 37 Sneden et al apj 222, 585, 1978
% 38 hi Li: why didnt mixing destroy Li?? Hartoog PASP 90, 167, 1978
% 39 Li isotope ratio Lambert \& Sawyer 1984
% he says sb freq is not unusual; see also Tomkin et al PASP 96, 609, 1984

By the mid-1980's several high-dispersion model-atmosphere abundance analyses of
CDRGs had been published (e.g., Sneden et al.\ 1978; Rao 1978; Lambert \& Sawyer
1984; and references therein). These studies confirmed that in representative
members of the class, carbon is deficient relative to iron by factors of about
10 to 30, nitrogen is enhanced by factors of up to 4, and metals have roughly
solar abundances, all relative to normal red giants. 

% Puzzles related to the abundance of lithium in these stars arose, and are
% discussed below.

In spite of the remarkable and poorly understood compositions of these stars,
some of them visible to the naked eye, there was a lull in follow-up studies for
a couple of decades. In the past few years, however, there has been a renewal of
interest, and several authors have published detailed analyses of their chemical
compositions and discussions of their evolutionary origins (e.g., Palacios et
al.\ 2012, 2016, hereafter P12 and P16; Adamczak \& Lambert 2013, hereafter
AL13).  

% Pourbaix?

\section{Str\"omgren Photometry}

\subsection{Carbon-Deficient Giants}

From 1974 to 1979 I carried out a program of photoelectric stellar photometry in
the \Stromgren\ {\it uvby\/} system, using 0.41-, 0.61-, and 0.91-m telescopes
at   Cerro Tololo Inter-American Observatory (CTIO) and Kitt Peak National
Observatory (KPNO)\null. My program was focused primarily on metal-deficient red
giants, many of which I had discovered during examination of objective-prism
photographic plates obtained with the Curtis Schmidt telescope at CTIO (and
earlier, when the telescope was located in Michigan). My final photometric
results for the metal-deficient stars were published almost four decades ago
(Bond 1980, hereafter B80). Details of the photometric reductions and
calibration to the standard {\it uvby\/} system are given in B80. 

In the course of these observations, I had also measured a selection of CDRGs,
but these data have remained unpublished. Because of recent renewed interest in
these stars (\S1), and the availability of new analysis tools and much more
precise parallaxes, I believe it is useful to present my results now.

Table~1 gives my photometric measurements of the CDRGs. Successive columns list
the star name, the visual magnitude $V$ (transformed from the $y$ magnitudes),
the $b-y$ color, and the color differences, defined as $m_1=(v-b)-(b-y)$ and
$c_1=(u-v)-(v-b)$. The fifth and sixth columns contain the number of nights on
which I made observations of each star and the Galactic latitude, and the final
column gives a reference to the first publication that reported the carbon
deficiency. The average uncertainties of a single observation for this ensemble,
calculated from the internal scatter for the stars observed more than once, are
$\pm$0.012, $\pm$0.006, $\pm$0.011, and $\pm$0.018 mag in $V$, $b-y$, $m_1$, and
$c_1$, respectively. Apart from the early discoveries recounted in \S1, most of
the listed stars were first recognized---many of them by the present writer---on
objective-prism plates taken in the Curtis Schmidt survey of the southern
hemisphere, and published by Bidelman \& MacConnell (1973, hereafter
BM73).\footnote{My photometry of the BM73 star BD\,$-$19~967, as well as its
recent \Gaia\/ parallax, indicate that it is a fairly normal G~dwarf, so it is
omitted from Table~1. Cottrell \& Norris (1978) had raised similar doubts about
the star, based on their photometry in the DDO system. The star may be a
misidentification by BM73 of the nearby BD\,$-$19~969, which I did not observe.
Of the remaining 33 CDRGs listed by BM73, I was able to obtain photometry of 31
of them, missing only HD\,119256 and HD\,124721.} Several more CDRGs were
discovered by myself, on Curtis Schmidt plates that I obtained during my
searches for metal-deficient stars as described in~B80. I have also included in
Table~1 {\it uvby\/} photometry for 37~Com, quoted from Crawford \& Perry
(1989).

Four out of the 41 stars listed in Table~1 (HR\,1229, HR\,4154, HR\,6757, and
HR\,6766) were measured by Eggen (1993) in a program of photometry on a modified
\Stromgren\ {\it uvby\/} system. His observations used a $v$ filter with a
different (narrower) bandpass than the standard filter, and he denoted his color
differences as $M_1$ and $C_1$. A comparison of our results, in the sense Bond
minus Eggen, gives these mean differences: $\Delta V=-0.016\pm0.003$,
$\Delta(b-y)=+0.006\pm0.002$, $m_1-M_1=0.000\pm0.004$, and
$c_1-C_1=-0.094\pm0.006$, with standard deviations of 0.007, 0.004, 0.007, and
0.012~mag, respectively. These differences indicate excellent agreement, with
small scatter, between our results, except for a large systematic offset between
$c_1$ and $C_1$. Nearly identical results were reported in B80, when I compared
my photometry of metal-deficient red giants with results published by Eggen in
several earlier papers.

% Not observed or omitted: BD -19 967 (dwarf/Id problem?); might actually be
% 		-19 969
% 	119256: only observed in 1970
% 	124721: observed in 1981
% 	188328: no obs card?? OH... it's a typo in BM73

\subsection{Barium Stars and Normal Giants}

Barium stars, or ``\BaII\ stars,'' are a class of peculiar late-type stars
showing enhanced abundances of carbon and of barium and other $s$-process
elements, which were first recognized by Bidelman \& Keenan (1951, hereafter
BK51). Barium stars posed a puzzle from the standpoint of stellar evolution,
since carbon and $s$-process elements were not expected to be created and
dredged up to the stellar surface until the asymptotic-giant-branch (AGB) stage;
however, it was known that \BaII\ stars only had approximately the luminosities
of normal red giants. The explanation came from the recognition that most or all
barium stars are members of wide spectroscopic binaries (e.g., McClure 1984),
indicating that they are companions of former AGB stars (now white dwarfs) which
have been contaminated on their surfaces by processed material accreted from a
stellar wind. For recent reviews of barium stars, see, for example, K\"appeler
et al.\ (2011), Escorza et al.\ (2017, 2019), Jorissen et al.\ (2019), and
references therein. 

The carbon enhancement in barium stars stands in sharp contrast to the carbon
deficiency of CDRGs. This was already noted by BK51, who stated that ``The
remarkable absence of CH in HR~885 thus represents a departure from a normal
spectrum in the opposite sense from that shown by the \BaII\ $\dots$ stars.''
Thus a comparison of the photometric properties of these two groups is of
interest. During the 1974--1979 observations described above, I also obtained
\Stromgren\ photometry of a large number of \BaII\ stars. These results are
presented in Table~2. It has the same format as Table~1, except that for
literature references for the individual barium stars, I refer to the articles
in the previous paragraph. In order to provide a sample of the most pronounced
barium stars, I only include in Table~2 the 41 stars from my observations that
have a ``barium index'' (Warner 1965) between Ba3 and Ba5 (the highest value),
according to the catalog of \BaII\ stars assembled by Escorza et al.\ (2017).

Finally, for comparison of these two peculiar groups with normal red giants, I
selected a sample of bright field stars of spectral type G0 and later, and
luminosity classes of II--III, from the catalog of {\it uvby\/} standard stars
published by Perry et al.\ (1987). There are 29 stars in that paper that satisfy
the selection criteria. Table~3 presents the {\it uvby\/} photometry for these
stars, and their spectral types, taken directly from Perry et al.

\section{The Anti-Bond-Neff Effect in Carbon-Deficient Giants}

Fifty years ago, John Neff and I published a paper (Bond \& Neff 1969, hereafter
BN69) on a surprising result we had found using intermediate-band photometry of
a small sample of barium stars. Our data showed that the \BaII\ stars'
spectral-energy distributions (SEDs) exhibit a broad absorption feature,
centered near 4000~\AA\null. In the prototypical \BaII\ star $\zeta$~Cap, the
absorption has a width of at least 1500~\AA\ and a maximum depth of about
0.3~mag, compared to the SED of a normal red giant of about the same spectral
type. Other authors have called this phenomenon the ``Bond-Neff effect''
(hereafter BNE), a terminology I immodestly adopt here.

The species responsible for the BNE was a puzzle for many years. BN69 noted that
the broad feature is similar to a pseudo-continuous absorption seen in
laboratory studies of the C$_3$ molecule (Brewer \& Engelke 1962; see also a
later paper by Snow \& Wells 1980), as well as in comets and very cool carbon
stars (see references in BN69). However, it would be surprising for C$_3$ to
exist in the atmospheres of stars as warm as the \BaII\ stars; moreover, its
presence was directly ruled out by Baird (1982), who showed that the expected
rotational lines of C$_3$ are absent in high-dispersion spectra of a bright
barium star. Williams (1975) suggested instead that the BNE is caused simply by
the increased strength of the known molecular bands of CN, CH, and C$_2$ in the
\BaII\ stars. However, Fix \& Neff (1975) and Fix (1976) used spectrophotometric
scans of several barium stars to show that the absorption feature is continuous
at the resolution of their data. In a conference abstract, McWilliam \& Smith
(1984) proposed alternatively that the BNE is due to the large number of
absorption lines of $s$-process rare-earth elements in the broad region around
4000~\AA\ in the spectra of barium stars.

% , but to my knowledge this study was never published in a journal paper.

The bandpass of the \Stromgren\ $v$ filter lies near the wavelength of the
maximum absorption of the BNE.\footnote{The $v$ filter has an effective wavelength of 4100~\AA\ and a width of 170~\AA\ (Bessell 2005).} Thus the $c_1$ index, which can be written
as $u -2v+b$, is very sensitive to the effect. I showed, in my study of
extremely metal-deficient giants (B80, Figure~7a), that the $c_1$ index is
highly correlated with the strength of the CH G~band, in stars with very weak
lines of other species. This finding would appear to rule out the
McWilliam--Smith suggestion as the dominant contributor, and it strongly pointed
to CH as the primary carrier of the BNE absorption feature.

Masseron et al.\ (2014, hereafter M14) now appear to have settled the issue. M14
performed an extensive update of the line list for the CH molecule, adding newly
identified energy levels, and including the role of broad predissociation
lines---which they showed are present in the spectrum of the Sun, but previously
unrecognized (see their Appendix~A for a discussion of the predissociation
phenomenon). By incorporating these new data into calculations of synthetic
stellar spectra for stars with enhanced carbon abundances, M14 were able to
reproduce the BNE absorption feature as originally presented by BN69. Most of
the BNE was indeed shown to be due to CH, but enhanced line blanketing due to
high $s$-process abundances in \BaII\ stars does also contribute.

Because of the strengthened CH absorption feature, \BaII\ stars have
systematically {\it lower\/} $c_1$ indices than normal red giants---which was
the observation leading to the BN69 discovery. Since the CDRGs are marked by an
{\it absence\/} of CH, they should exhibit an {\it anti}-BNE; i.e., they would
be expected to have {\it higher\/} $c_1$ indices than normal red giants. I will
now show that this is indeed the case, by intercomparing the \Stromgren\
photometry of the CDRGs, normal red giants, and \BaII\ stars presented above.

First, however, the photometry listed in Tables~1, 2, and 3 has to be corrected for
interstellar extinction. Estimation of extinction for nearby stars has become much easier
in the past few years because of two developments: first, an online
tool\footnote{\url{https://stilism.obspm.fr/}} is available for estimating
reddening, $E(B-V)$, at any given Galactic position and distance (Capitanio et
al.\ 2017); second, precise stellar distances are now available from the recent
\Gaia\/ Data Release~2 (DR2; Gaia Collaboration et al.\ 2018).

Using these two tools, I determined the reddening for each of the stars in
Tables~1, 2, and 3. Table~4 presents the \Gaia\/ DR2 parallax, the \Gaia\/
photometry (apparent magnitude $G$, color index $BP-RP$), and the reddening for the CDRGs, barium stars, and field red giants. (A few of the red
giants in Table~3 are so bright that they are not included in DR2, and are
omitted from Table~4.) In determining the distances, I adjusted each \Gaia\/
parallax upward by 0.029~mas, as recommended by Lindegren et al.\ (2018). In
most cases, the reddening values are low, which is not surprising since the
stars are relatively bright and nearby; however, a few of them have higher
values. Based on the reddenings given in Table~4, I then corrected the
\Stromgren\ photometry in Tables~1, 2, and 3, using the formulae given by Crawford
(1975): $E(b-y)/E(B-V) = 0.74$, $E(m_1)/E(b-y) = -0.32$, and $E(c_1)/E(b-y) =
0.20$.

Figure~1 plots the dereddened $c_1$ color difference, denoted $(c_1)_0$, versus
the dereddened color index, $(b-y)_0$. The CDRGs are plotted as filled blue
circles, the normal giants as green filled triangles, and the \BaII\ stars as red
filled diamonds. The figure dramatically illustrates the BNE: {\it all\/} of
the barium stars have {\it smaller\/} values of $(c_1)_0$ than the field red
giants. The offset to lower $c_1$ values increases with redder $(b-y)_0$ color
(cooler temperatures), as the strength of the CH absorption increases. The
effect on the $c_1$ index reaches values as large as $\sim$0.8~mag for the most
extreme cases.

The CDRGs in Figure~1 behave oppositely, as expected based on the BNE being due
primarily to CH absorption, and because these stars lack CH in their
atmospheres. In nearly every case, their $c_1$ indices are {\it higher\/} than
those of normal red giants.\footnote{A similar finding regarding the $c_1$
indices of CDRGs was reported in a conference abstract by Herr \& MacConnell
(1972), but without details. Cottrell \& Norris (1978) obtained photometry of
several CDRGs in the DDO intermediate-band system, and noted that they are bluer
in the $C(42-45)$ color than normal red giants; they attributed this to the
absence of CH absorption in the 4200~\AA\ bandpass. Eggen (1993) demonstrated
that his $C_1$ index is high in several carbon-deficient stars, compared to
normal red giants and \BaII\ stars, but he interpreted the result as indicating
excess ``emission'' around 4000~\AA\ in the former.} The few cases of $c_1$
values mixed with the red giants may indicate less extreme degrees of carbon
deficiency.\footnote{AL13 remark briefly that the star HD~121071 is an example
of a CDRG with an intermediate C abundance; however, this designation arises
from a typographical error in Hartoog (1978), with the correct carbon-deficient
star being a different object, HD~121070.}

\begin{figure}[ht]
\centering
\includegraphics[scale=.6]{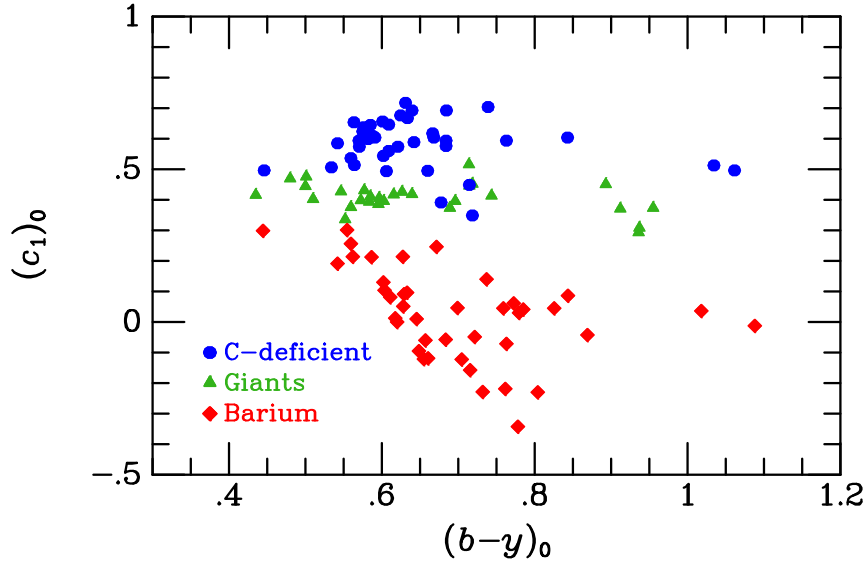}
% \plotone{c1_vs_b-y.eps}
\caption{
Reddening-corrected color difference $(c_1)_0$ vs.\ corrected $(b-y)_0$ color
index for carbon-deficient red giants ({\it blue filled circles}), normal field
red giants ({\it green filled triangles}), and barium stars ({\it red filled
diamonds}). Uncertainties are generally smaller than the plotting symbols.
The barium stars have systematically low $c_1$ indices, due to the Bond-Neff
absorption around 4000~\AA\ produced by much stronger CH features than in the
normal giants (see text). In contrast, the C-deficient red giants, lacking the
CH absorption, have generally higher $c_1$ values than the normal giants---and thus show
an ``anti-Bond-Neff effect.''
}
\end{figure}

The BNE and anti-BNE have several important implications. The CH absorption
behaves like a continuous opacity, which is strong in \BaII\ stars, but still
present in normal stars such as the field red giants in Figure~1, and even in
the Sun. At least until the recent work of M14, this ``missing'' opacity has not
been included in stellar-atmosphere modeling. One result is that, when CH is
present, metallic absorption lines will be formed higher in the atmosphere in
the region around 4000~\AA, and thus weakened. Abundance determinations will
therefore give systematically low results for lines in this spectral region, if
the model atmosphere does not include the CH opacity. Luck \& Bond (1982, their
Figure~1) demonstrated that this was the case in three subgiant CH stars (or
dwarf barium stars), where metallic lines below $\sim$4200~\AA\ gave lower
abundances than lines at longer wavelengths.

\begin{figure}[ht]
\centering
\includegraphics[scale=.6]{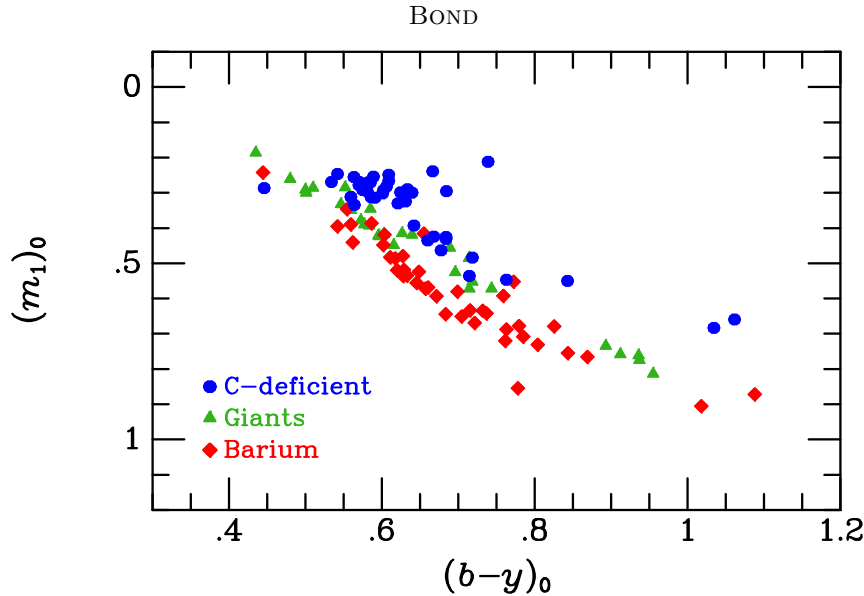}
% \plotone{c1_vs_b-y.eps}
\caption{
Reddening-corrected color difference $(m_1)_0$ vs.\ corrected $(b-y)_0$ color
index for carbon-deficient red giants ({\it blue filled circles}), normal field
red giants ({\it green filled triangles}), and barium stars ({\it red filled
diamonds}). As in Figure~1, uncertainties are generally smaller than the plotting
symbols. The barium stars have systematically high $m_1$ indices, due to the
enhanced Bond-Neff absorption in the $v$ band. Carbon-deficient red giants have
lower $m_1$ values than the normal giants, because of the absence of CH.
}
\end{figure}

Figure~2 shows the reddening-corrected metallicity index, $(m_1)_0$, plotted
against the corrected $(b-y)_0$ color. Recalling the definition,
$m_1=(v-b)-(b-y)$, we see that the $m_1$ index will also be affected by the
Bond-Neff absorption in the $v$ bandpass, but by a lesser amount than the $c_1$
index. The figure shows that the $m_1$ index is indeed higher in \BaII\ stars
than in normal red giants, but the effect is not as large as in Figure~1, since
the $v$ magnitude is not multiplied by two. The CDRGs have slightly lower $m_1$
indices than the red giants and barium stars, again due to the lack of the CH
absorption feature around 4000~\AA\null. 

The \Stromgren\ $m_1$ index is usually a useful indicator of metal content
(e.g., B80 and references therein), but Figure~2 shows that it is systematically
affected in stars with anomalous carbon content, independently of the abundance
of heavier elements.

\section{Color-Magnitude Diagrams and Initial Masses}

The recent availability of \Gaia\/ DR2 parallaxes allows considerably more
precise placement of nearby stars in the color-magnitude diagram (CMD)\footnote{ Throughout this paper, ``CMD'' means a plot of {\it absolute\/} magnitude vs.\ color.} than
was possible with astrometry from previous sources such as \Hipp. I will present
CMDs for the CDRGs, barium stars, and field red giants by plotting the
extinction-corrected \Gaia\/ absolute magnitude $(M_G)_0$ versus corrected
\Gaia\/ color index $(BP-RP)_0$. I made extinction corrections by using the
$E(B-V)$ values given in Table~4, and applying the approximate relations
$A_G\simeq2E(B-V)$ and $E(BP-RP)\simeq E(B-V)$ from Andrae et al.\ (2018). I
determined the absolute magnitudes based on the DR2 parallaxes also in Table~4,
with the Lindegren et al.\ (2018) correction applied.\footnote{I use the
broad-band \Gaia\/ photometry, rather than $M_V$ vs.\ $(b-y)$ based on my data
in this paper, because 
it is formally more precise
than my ground-based photometry, is all-sky, from space, with a single
instrument and filters, and should be less affected by the Bond-Neff feature.}

The resulting CMDs are shown in Figure~3 (left) for the \BaII\ stars and field
red giants (color-coded as in Figures~1 and~2), and in Figure~3 (right) for the
CDRGs. Superposed on both figures are evolutionary tracks in the \Gaia\/
photometric system, for single stars of masses 1.0 to $4.5\,M_\odot$, in steps
of $0.5\,M_\odot$, obtained from the MESA Isochrones and Stellar Tracks (MIST,
version 1.2; Dotter 2016; Choi et al.\ 2016)
website\footnote{\url{http://waps.cfa.harvard.edu/MIST/}} and its web
interpolator. These tracks assume an initial rotation of $v/v_{\rm crit} = 0.4$.
According to recent abundance analyses (e.g.,
AL13 and P16), CDRGs are typically slightly metal-deficient, with an average
$\rm[Fe/H]\simeq-0.2$ to $-0.3$. The tracks plotted in Figure~3 assume $\rm[Fe/H]=-0.25$.

\begin{figure}[ht]
\centering
\includegraphics[scale=.525]{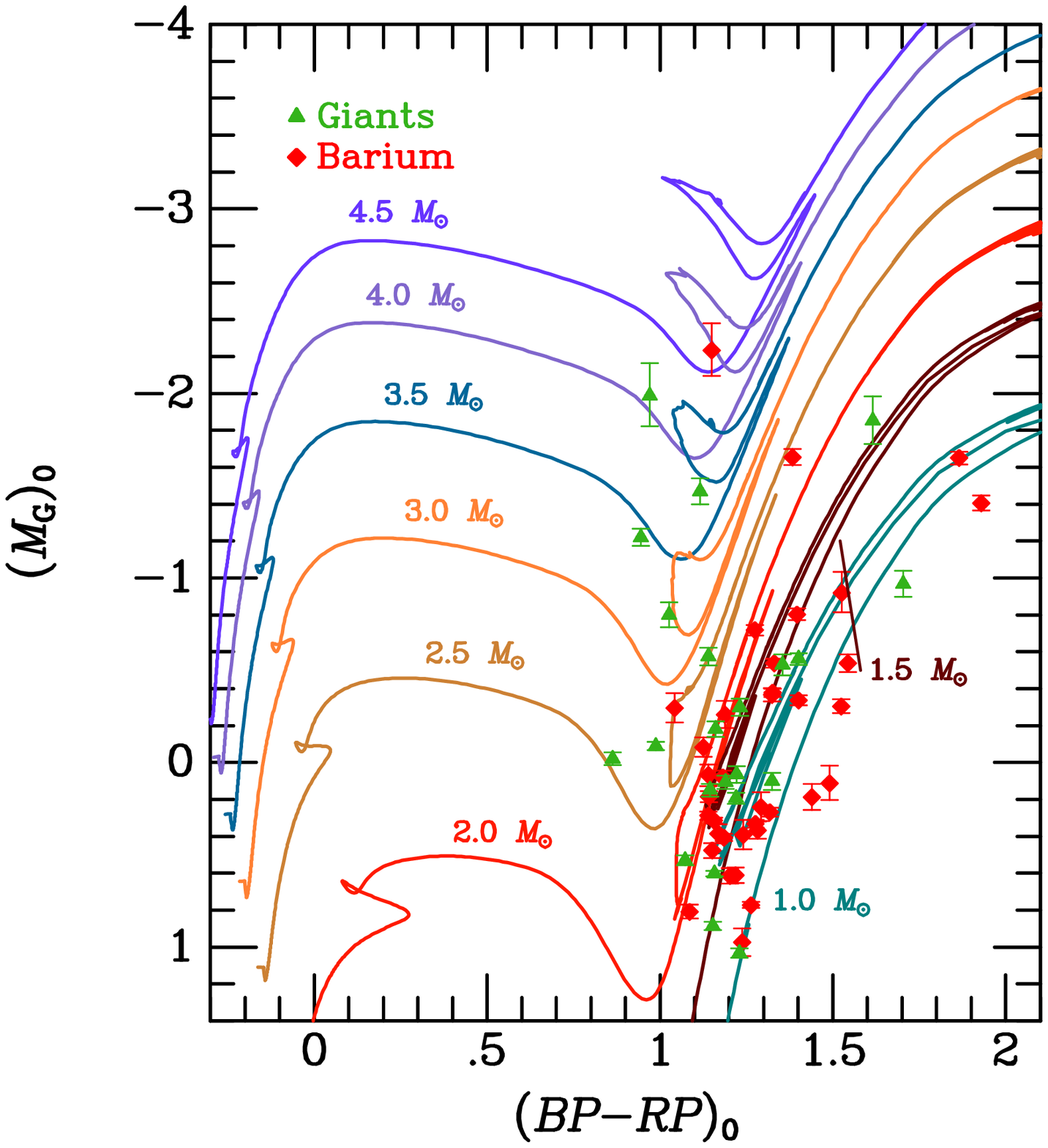}
\hskip0.3in
\includegraphics[scale=.525]{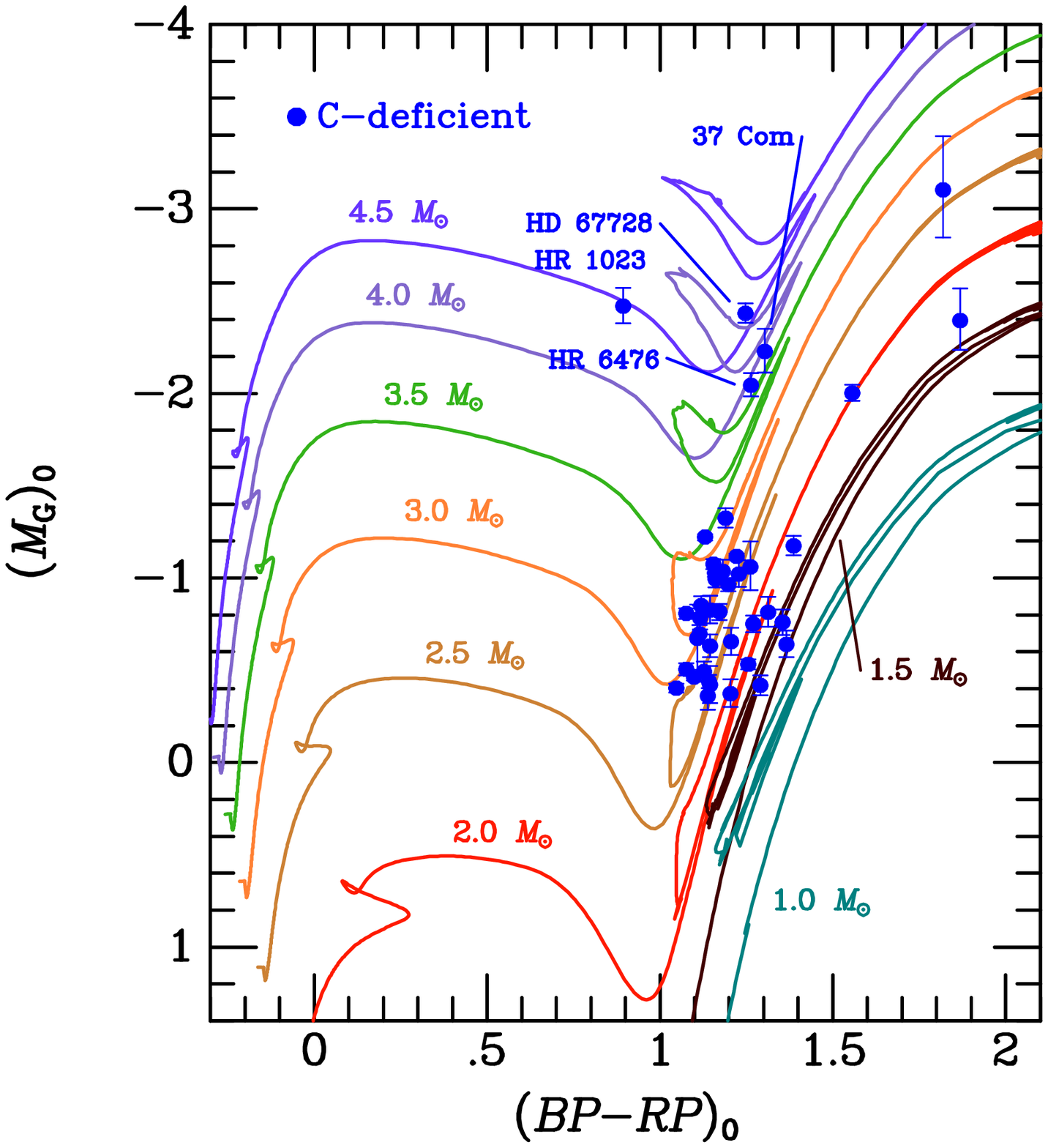}
\caption{
Color-magnitude diagrams for barium stars and field red giants (left panel) and for
carbon-deficient red giants (right panel). Plotting symbols are the same as in Figures 1 and 2. Extinction-corrected absolute
magnitudes are plotted against corrected color indices in the \Gaia\/
photometric system. Superposed on the data are theoretical stellar evolutionary
tracks for stars of masses 1.0 to $4.5\,M_\odot$, and metallicity $\rm[Fe/H]=-0.25$, obtained
using the MIST web tool (see text). The barium stars and field giants sample a
wide range of initial masses and evolutionary stages. In contrast, most of the
carbon-deficient red giants fall into a tight clump, suggesting that most of
them have initial masses of about 2 to $3.5\,M_\odot$---if they have evolved as
single stars. Four exceptions at apparently higher initial masses are marked: HR~1023,
HD~67728, 37~Com, and HR~6476 (see text for discussion).
}
\end{figure}

The barium stars and field red giants in Figure~3 (left) show a wide range in
CMD location and the implied initial stellar masses. This is not surprising,
since the \BaII\ stars are considered to be ``bystanders'' that were polluted by
a former AGB companion. Thus they should be drawn at random from the population
of normal red giants, as appears to be the case in the figure.

By contrast, the CDRGs have a very different distribution in the CMD\null. Most
of them are concentrated in a fairly tight clump, implying that most of them had
initial masses of about 2 to $3.5\,M_\odot$---on the assumption that they have
evolved as single stars. There are a few outliers of initial masses as low as
$\sim\!1.5\,M_\odot$, and three luminous stars that appear to have evolved to
higher luminosity from the main clump. The requirement of a fairly sharp
minimum  mass for stars that exhibit the carbon deficiency appears to be a
robust result, since the CDRGs were selected in spectroscopic surveys without
regard to, or even advance knowledge of, their absolute magnitudes.

Apart from the clumped stars, there are four luminous outliers, all lying near
the single-star tracks for stars of $\sim\!4$ to $4.5\,M_\odot$. They are labelled in
Figure~3 (right): HR~1023, HD~67728, 37~Com, and HR~6476. It is striking that
AL13, in their high-resolution spectroscopic analysis of 24 CDRGs, singled out
HR~1023, HD~67728, and 37~Com as having unusually high rotational velocities
compared to the rest of their sample: these stars' values of $v\sin i$ are 22.7
(HR~1023; de~Medeiros \& Mayor 1999), 13.0 (HD~67728; P16), and $11.0\,\kms$
(37~Com; Drake et al.\ 2002). The AL13 sample did not include HR~6476, but its
rotation is also unusually high compared to normal red giants, with $v\sin
i=7.3\,\kms$, according to Hekker \& Mel{\'e}ndez (2007).

My conclusion that most of the CDRGs had initial masses of $\sim$2 to
$3.5\,M_\odot$, with a fairly sharp lower-mass cutoff, generally agrees with
earlier published results, but strengthens and refines them because of the
higher precision of the \Gaia\/ parallaxes. For example, based on locations in
the CMD determined from \Hipp\/ parallaxes, AL13 inferred that their sample of
two dozen CDRGs have masses in the range $\sim$2.5 to $5\,M_\odot$, with a mean
of about $3.6\,M_\odot$. P16, also using \Hipp\/ distances, refined the mass
range for their sample of 19 CDRGs to 3.2 to $4.2\,M_\odot$. The existence of a
distinct smaller group of CDRGs with higher masses of about 4 to $4.5\,M_\odot$ and
unusually high rotations is, to my knowledge, a new result.

\section{Evolutionary Status}

Depletion of carbon is a strong signature of material that has been exposed to
hydrogen-burning via the CN-cycle. When equilibrium is reached, C is
underabundant by a factor of about 20, the carbon isotopic ratio \Cratio\
declines to about 3, and the abundance of $^{14}$N is increased, such that the
sum of $^{12}$C, $^{13}$C, and $^{14}$N remains constant (e.g., Iben 1967; AL13). The abundance of
lithium is drastically reduced. In general, published CNO abundance analyses of
CDRGs give results in agreement with these expectations (see P12, AL13, P16, and
references therein), except that some GDRGs are not depleted in Li. These
findings indicate that the surfaces of these stars are strongly contaminated with
material that was once deep in the hydrogen-burning core. 

The papers by P12, AL13, and P16 give extended discussions of possible
evolutionary scenarios to explain the existence of these rare stars. They
generally reach the conclusion that ``the weak-G-band puzzle [is] largely
unsolved'' (P16). As P12 and P16 point out, their locations in the CMD (e.g., my
Figure~3 right) are consistent with the stars being either (1)~subgiants that
have just reached the base of the red-giant branch (RGB)---which is about the
location of the onset of the first ``dredge-up'' in normal red giants, or
(2)~stars that reached the tip of the RGB, ignited core helium-burning, and
dropped back to the red-giant clump. But in neither case does standard
evolutionary theory predict a surface composition dominated by fully
CN-processed material. AL13 discuss the key finding, reinforced by my results
here, that CDRGs are more massive on average than normal field red giants. They
raise the possibility that rotational mixing, due to the faster rotations of
main-sequence stars in this mass range as compared to the lower-mass progenitors
of typical red giants, could be the cause of the mixing to the surface. This
scenario to explain CDRGs has recently been discussed by Smiljanic et al.\
(2018). However, many main-sequence stars in this mass range are fast rotators,
but CDRGs are extremely rare. On the other hand, the fast rotations associated
with the four most massive CDRGs that I have noted above may be an important new
clue.

The above discussions have considered the evolution of single stars. In a recent
paper, Izzard et al.\ (2018) discuss mass-transfer and stellar mergers, in the
context of explaining the presence of fairly massive stars in the old population
of the Galactic thick disk. They raise the possibility that C-poor and N-rich
stars could be the result of binary-star interactions that expose CN-processed
material at the surface, such as in the CDRGs. 

A striking feature of the data presented in Table~1 is that many of the CDRGs
lie at fairly high Galactic latitudes, in spite of them appearing to be stars of
higher masses than normal red giants. Note that the BM73 survey covered
essentially the entire southern hemisphere, so there should not be a bias against low latitudes. 

Is it possible that CDRGs are descended from binary systems that have merged or
undergone mass-accretion, giving them enhanced masses compared to the single stars that
are currently evolving in their host population? In this case, they should mimic the properties of an older population, compared to single stars of the same mass. I performed an experiment of extracting
from \Gaia\/ DR2 a sample of normal red giants that simulates the BM73
selection. I required the DR2 stars to lie in the southern hemisphere as do most
of the BM73 stars ($\delta<0^\circ$), to be brighter than the approximate
magnitude limit of the BM73 sample ($G<8.9$), and to lie in the main clump of
CDRGs in the CMD shown in Figure~3 (right), i.e., $1.05 \le(BP-RP) \le 1.4$ and
$-0.3 \ge M_G \ge -1.3$. This selection resulted in 1,883 stars. The BM73
sample\footnote{To avoid introducing a bias, I did not include the five stars in
Table~1 that I had discovered, because my objective-prism survey was limited to
high-latitude fields.} contains 28 CDRGs lying within the same box in the CMD,
showing that they represent only 1.5\% of the red giants, even when the red
giants are limited to this relatively narrow range of colors and absolute
magnitudes. I then calculated the absolute value of the distance from the
Galactic plane, $|Z|$, for each star in the two samples, using their known
distances and Galactic latitudes.

Figure~4 shows the cumulative distribution of $|Z|$ values for the field giants
and CDRGs. It illustrates that the CDRGs indeed do systematically lie at larger
distances from the Galactic plane than the field red giants. This could be
consistent with a scenario in which the CDRGs have arisen from an older and
dynamically hotter population than the field giants, and have acquired their
higher masses through relatively recent binary mergers or mass transfer. A Kolmogorov-Smirnov test
indicates a probability of only $\sim$1.7\% that the CDRG distribution of $|Z|$
values in Figure~4 was drawn from the red-giant population.

\begin{figure}[ht]
\centering
\includegraphics[scale=.5]{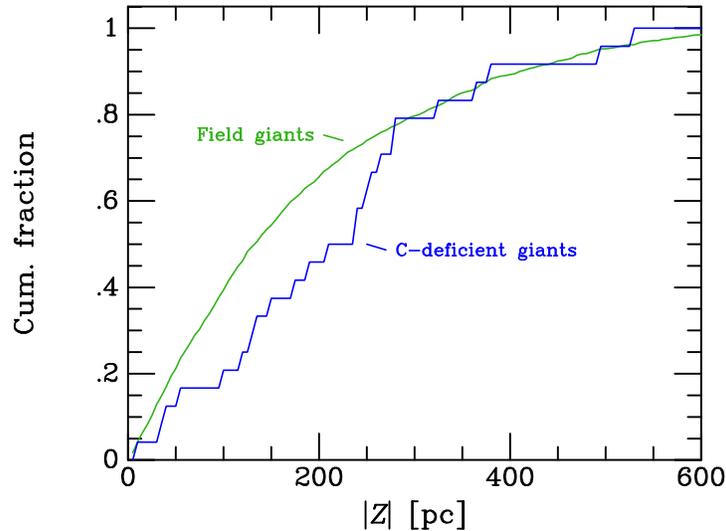}
\caption{
Cumulative distributions of the absolute value of the distance from the Galactic
plane, $|Z|$, for the BM73 sample of CDRGs lying in the main clump of stars in
Figure~3 (right), shown as a blue line, and for a sample of field
southern-hemisphere red giants lying in the same CMD clump selected from \Gaia\/
DR2 with the same magnitude limit as the CDRGs (green line). As discussed in the
text, the CDRGs are distributed to systematically higher $|Z|$ distances,
suggesting that they may arise from older and lower-mass progenitors that have
undergone binary interactions and mass augmentation.
}
\end{figure}

\section{Summary and Future Work}

CDRGs are a rare class of red giants whose atmospheres are composed of material
exposed to CN-cycle hydrogen-burning. The main conclusions of this study are the
following:

1.~I present \Stromgren\ {\it uvby\/} photometry for a nearly complete sample of
the known CDRGs, and for a selection of barium stars and normal red giants.
Barium stars exhibit the ``Bond-Neff effect'' of unusually low  \Stromgren\
$c_1$ indices, now known to be caused by a broad CH absorption feature centered
around 4000~\AA\null. The CDRGs, lacking CH absorption, show an ``anti-Bond-Neff
effect'' of unusually high $c_1$ indices compared to normal red giants.

2.~The locations of CDRGs in the color-absolute magnitude diagram are now well defined because of
precise parallaxes from \Gaia\/ DR2. Most of them lie in a tight clump in the
CMD consistent with initial masses of about 2 to $3.5\,M_\odot$. A second,
smaller, and possibly distinct group has higher masses of $\sim\!4$ to $4.5\,M_\odot$,
and they all exhibit unusually high rotational velocities.

3.~The evolutionary status of CDRGs remains unclear, as does the mechanism that
has brought the highly processed material to the surface. It is uncertain
whether CDRGs are hydrogen-burning stars that have just reached the bottom of
the red-giant branch and the onset of the first dredge-up, or are more highly
evolved helium-burning red-giant clump stars. Another mystery is why there is an
apparent lower-mass cutoff for the occurrence of the carbon-depletion
phenomenon.

4.~A hint that CDRGs might be members of binaries that have increased in mass
through mass-transfer or mergers comes from their systematically higher
distances from the Galactic plane than normal red giants lying in the same
location in the CMD\null. The high rotation of the high-luminosity subset may
support such speculation.

Nearly seven decades after Bidelman's discovery of the peculiarities of HR~885,
these stars continue to pose one of the most poorly understood puzzles in
stellar evolution. There are several avenues for future studies. A systematic
test of binarity would be extremely useful. As many authors have noted, the
sample of known CDRGs remains distressingly small, a little over three dozen;
thus spectroscopic surveys that would find more of them are highly desirable.
One project that could approximately double the known number would be a
systematic examination of an existing collection of objective-prism plates
covering the entire northern sky, obtained with the Burrell Schmidt telescope at
KPNO (Bidelman 1998; see also Bond 2017).

\acknowledgments

My interest in stars with peculiar spectra was inspired many years ago by my
teacher, W.~P.~Bidelman. I thank the staffs of CTIO and KPNO for generous
support during my observing runs of several decades ago. Partial support of my
research at that time came from the National Science Foundation in a grant to
Louisiana State University (AST 78-25538).

This work has made use of data from the European Space Agency (ESA) mission {\it
Gaia\/} (\url{https://www.cosmos.esa.int/gaia}), processed by the {\it Gaia\/}
Data Processing and Analysis Consortium (DPAC,
\url{https://www.cosmos.esa.int/web/gaia/dpac/consortium}). Funding for the DPAC
has been provided by national institutions, in particular the institutions
participating in the {\it Gaia\/} Multilateral Agreement.

This research has made use of the SIMBAD database and the VizieR catalogue access tool, CDS, Strasbourg, France.

% Plot V-G vs color? will C-def \& Ba II be different??

% put in the ones I never got photometry for?  The other Bidelman-MacConnell
% stars??

% \end{document}

\clearpage

\begin{deluxetable}{lccccccl}
% \tabletypesize{\footnotesize}
\tablecaption{\Stromgren\ Photometry of Carbon-Deficient Red Giants}
\tablewidth{0pt}
\tablehead{
\colhead{Star} & 
\colhead{$V$} & 
\colhead{$b-y$} & 
\colhead{$m_1$} & 
\colhead{$c_1$} &
\colhead{$n$\tablenotemark{a}} &
\colhead{$b$ [deg]} &
\colhead{Discovery\tablenotemark{b}}
}										 	
\startdata									 	  
CD $-$28 75   & 8.967  &   1.039 &    0.682 &    0.514 &        1 & $-$82.65 & HEB  \\      
% HD 13763      & 6.403  &   0.878 &    0.687 &    0.381 &        3 $-$63.43 & & HEB  \\    
HD 17232      & 8.723  &   1.080 &    0.653 &    0.500 &        2 & $-$58.75 & HEB   \\     
HR 885        & 5.475  &   0.574 &    0.236 &    0.591 &        4 & $-$10.21 & C12,B51  \\  
HD 18636      & 7.645  &   0.571 &    0.253 &    0.655 &        3 & $-$61.27 & BM73  \\     
HD 20090      & 8.052  &   0.782 &    0.541 &    0.597 &        3 & $-$56.02 & HEB   \\     
HR 1023       & 6.364  &   0.549 &    0.254 &    0.517 &        5 & $-$41.25 & BM73   \\    
BD +5 593     & 9.361  &   0.835 &    0.185 &    0.651 &        4 & $-$31.91 & BM73   \\    
% HD 26549    & 6.983  &   0.910 &    0.713 &    0.400 &        3 & 	     &	      
HR 1299       & 6.451  &   0.665 &    0.404 &    0.496 &        2 & $-$47.05 & BM73   \\    
HD 28932      & 7.942  &   0.643 &    0.271 &    0.501 &        4 & $-$29.70 & BM73  \\     
% BD -19 967  & 9.881  &   0.407 &    0.203 &    0.365 &        4 & 	     & BM73   \\  
HD 30297      & 8.575  &   0.712 &    0.216 &    0.581 &        4 & $+$03.00 & B57  \\      
HD 31274      & 7.121  &   0.607 &    0.300 &    0.657 &        3 & $-$39.46 & BM73   \\    
HD 31869      & 9.286  &   0.578 &    0.276 &    0.595 &        3 & $-$35.93 & BM73   \\    
HD 36552      & 8.077  &   0.542 &    0.267 &    0.508 &        3 & $-$32.54 & BM73   \\    
HD 40402      & 8.596  &   0.581 &    0.266 &    0.576 &        2 & $-$23.22 & BM73   \\    
HD 49960      & 8.346  &   0.657 &    0.288 &    0.683 &        2 & $-$14.03 & BM73  \\     
HD 54627      & 8.775  &   0.621 &    0.302 &    0.618 &        2 & $-$16.81 & BM73   \\    
HD 56438      & 8.091  &   0.674 &    0.277 &    0.676 &        2 & $-$15.84 & BM73   \\    
HD 67728      & 7.542  &   0.713 &    0.452 &    0.398 &        2 & $+$07.00 & BM73  \\     
HD 78146      & 8.571  &   0.734 &    0.403 &    0.617 &        2 & $+$12.68 & BM73   \\    
HD 82595      & 8.189  &   0.666 &    0.314 &    0.725 &        2 & $+$10.68 & BM73   \\    
HD 91622      & 8.224  &   0.703 &    0.419 &    0.598 &        3 & $+$52.49 & HEB  \\      
HR 4154       & 6.108  &   0.602 &    0.265 &    0.649 &        2 & $+$12.62 & BM73   \\    
HD 94956      & 8.460  &   0.609 &    0.281 &    0.630 &        2 & $+$27.33 & BM73   \\    
HD 102851     & 8.788  &   0.684 &    0.286 &    0.701 &        2 & $+$10.25 & BM73   \\    
CD $-$37 7613 & 9.842  &   0.624 &    0.283 &    0.640 &        1 & $+$23.69 & BM73   \\    
HD 105783     & 9.092  &   0.793 &    0.195 &    0.714 &        1 & $+$20.20 & BM73   \\    
37 Com\tablenotemark{c} & 4.890 & 0.726 & 0.482 & 0.350&        4 & $+$85.86 & R52   \\
HD 120170     & 9.041  &   0.604 &    0.263 &    0.603 &        3 & $+$51.55 & BM73  \\     
HR 5188       & 5.940  &   0.940 &    0.519 &    0.623 &        1 & $-$20.06 & BM73   \\    
HD 132776     & 8.827  &   0.740 &    0.413 &    0.588 &        3 & $+$46.69 & BM73   \\    
HD 146116     & 7.697  &   0.700 &    0.305 &    0.589 &        4 & $+$34.00 & BM73   \\    
HR 6476       & 5.758  &   0.807 &    0.506 &    0.467 &        3 & $+$23.43 & ST69  \\     
HR 6757       & 6.338  &   0.698 &    0.262 &    0.563 &        4 & $+$09.86 & BM73   \\    
HR 6766       & 4.559  &   0.614 &    0.265 &    0.647 &        3 & $-$04.02 & BM73   \\    
HR 6791       & 5.006  &   0.578 &    0.330 &    0.516 &        5 & $+$25.90 & GK58  \\     
HD 188028\tablenotemark{d} & 7.741 &  0.665 &  0.290 &  0.619 & 2 & $-$24.82 & BM73    \\   
HD 198718     & 8.631  &   0.592 &    0.274 &    0.641 &        3 & $-$39.97 & BM73   \\    
HD 201557     & 9.247  &   0.656 &    0.388 &    0.592 &        3 & $-$41.16 & BM73   \\    
HD 204046     & 9.009  &   0.721 &    0.284 &    0.700 &        2 & $-$45.87 & BM73  \\     
HD 207774     & 8.929  &   0.590 &    0.302 &    0.542 &        2 & $-$43.09 & BM73   \\    
% HD 210043     & 7.657  &   0.850 &    0.579 &    0.308 &        2 $-$45.65 & &   \\	    
HD 215974     & 8.492  &   0.629 &    0.241 &    0.616 &        3 & $-$29.21 & HEB  \\      
\enddata
\tablenotetext{a}{Number of photometric observations.}
\tablenotetext{b}{Reference for first discovery of carbon-deficient nature.
Reference codes are: B51 (Bidelman 1951); B57 (Bidelman 1957); BM73 (Bidelman \&
MacConnell 1973); C12 (Cannon 1912); HEB (this paper; discovered by author on
Curtis Schmidt plate); R52 (Roman 1952); ST69 (Spinrad \& Taylor 1969).}
\tablenotetext{c}{Photometry quoted from Crawford \& Perry (1989).}
\tablenotetext{d}{Due to a typographical error, Bidelman \& MacConnell (1973)
designated this star as HD\,188328.}
\end{deluxetable}

% \end{document}

\begin{deluxetable}{lccccc}
% \tabletypesize{\footnotesize}
\tablecaption{\Stromgren\ Photometry of Barium Red Giants}
\tablewidth{0pt}
\tablehead{
\colhead{Star} & 
\colhead{$V$} & 
\colhead{$b-y$} & 
\colhead{$m_1$} & 
\colhead{$c_1$} &
\colhead{$n$\tablenotemark{a}} 
}
\startdata
HD 4084       & 8.611 &     0.679 &	0.591 &           0.248  &  1 \\
HD 4395       & 7.672 &     0.446 &	0.242 &           0.299  &  3 \\
HR 774        & 5.794 &     0.771 &	0.685 & \llap{$-$}0.069  &  3 \\
HD 19014      & 8.215 &     1.080 &	0.885 &           0.049  &  2 \\
HD 20394      & 8.734 &     0.679 &	0.461 &           0.095  &  3 \\
HD 24035      & 8.510 &     0.763 &	0.625 & \llap{$-$}0.222  &  2 \\
HD 26886      & 7.992 &     0.587 &	0.336 &           0.308  &  3 \\
HD 29370      & 9.312 &     0.653 &	0.553 &           0.012  &  2 \\
HD 31487      & 8.094 &     0.843 &	0.530 &           0.075  &  3 \\      
HD 32712      & 8.527 &     0.708 &	0.578 &           0.048  &  2 \\
HD 36598      & 8.030 &     0.782 &	0.713 & \llap{$-$}0.214  &  2 \\
HD 43389      & 8.330 &     0.920 &	0.749 & \llap{$-$}0.032  &  2 \\
HR 2392       & 6.284 &     0.659 &	0.571 & \llap{$-$}0.059  &  3 \\
HD 49641      & 7.148 &     0.800 &	0.671 &           0.034  &  3 \\
HD 50082      & 7.455 &     0.610 &	0.446 &           0.132  &  2 \\
HD 62017      & 8.814 &     0.564 &	0.388 &           0.196  &  1 \\
HD 65854      & 8.419 &     0.571 &	0.438 &           0.215  &  3 \\
HD 82221      & 7.960 &     0.810 &	0.700 &           0.046  &  1 \\
HD 84678      & 8.984 &     0.929 &	0.806 & \llap{$-$}0.312  &  1 \\
HD 88035      & 9.139 &     0.656 &	0.527 &           0.101  &  1 \\
HD 88562      & 8.550 &     0.874 &	0.745 &           0.092  &  2 \\
HD 89175      & 7.720 &     0.681 &	0.514 & \llap{$-$}0.089  &  1 \\
HD 91208      & 8.052 &     0.580 &	0.383 &           0.260  &  2 \\
HD 92626      & 7.142 &     0.824 &	0.725 & \llap{$-$}0.226  &  1 \\
HR 4474       & 6.137 &     0.637 &	0.516 &           0.093  &  4 \\
HD 107541     & 9.364 &     0.673 &	0.410 & \llap{$-$}0.118  &  1 \\
HR 5058       & 5.115 &     0.708 &	0.650 & \llap{$-$}0.122  &  2 \\
HD 120620     & 9.628 &     0.645 &	0.531 &           0.055  &  1 \\
HD 123949     & 8.765 &     0.862 &	0.667 &           0.052  &  1 \\
HD 178717     & 7.163 &     1.208 &	0.833 &           0.011  &  2 \\
HD 183915     & 7.315 &     0.821 &	0.572 &           0.057  &  3 \\
HD 199435     & 8.305 &     0.667 &	0.470 &           0.022  &  2 \\
HD 199939     & 7.422 &     0.749 &	0.660 & \llap{$-$}0.043  &  2 \\
HD 201657     & 8.022 &     0.734 &	0.628 & \llap{$-$}0.047  &  2 \\
HD 201824     & 8.964 &     0.643 &	0.512 &           0.005  &  2 \\
$\zeta$ Cap   & 3.753 &     0.607 &	0.418 &           0.105  &  1 \\
HD 204886     & 8.169 &     0.749 &	0.638 &           0.142  &  1 \\
HD 205011     & 6.439 &     0.641 &	0.475 &           0.216  &  1 \\
HD 211594     & 8.094 &     0.696 &	0.557 & \llap{$-$}0.112  &  3 \\
HD 219116     & 9.286 &     0.599 &	0.382 &           0.215  &  2 \\
CPD $-$64 4333& 9.618 &     0.725 &	0.630 & \llap{$-$}0.155  &  1 \\
\enddata
\tablenotetext{a}{Number of photometric observations.}
\end{deluxetable}

% \end{document}

\begin{deluxetable}{lccccl}
% \tabletypesize{\footnotesize}
\tablecaption{\Stromgren\ Photometry of Normal Field Red 
  Giants\tablenotemark{a}}
\tablewidth{0pt}
\tablehead{
\colhead{Star} & 
\colhead{$V$} & 
\colhead{$b-y$} & 
\colhead{$m_1$} & 
\colhead{$c_1$} &
\colhead{Sp.~Type} 
}
\startdata
HR 373      &   5.411 &  0.554 &  0.285 &  0.335  &   G5 IIIe	  \\
HR 617      &   2.000 &  0.696 &  0.526 &  0.395  &   K2 IIIab	  \\
HR 1030     &   3.613 &  0.547 &  0.333 &  0.426  &   G6 III	  \\
HR 1327     &   5.262 &  0.513 &  0.286 &  0.402  &   G5 IIb	  \\
HR 1346     &   3.637 &  0.596 &  0.422 &  0.385  &   K0 IIIab	  \\
HR 1373     &   3.759 &  0.597 &  0.424 &  0.405  &   K0 III	  \\
HR 1409     &   3.529 &  0.616 &  0.449 &  0.417  &   G9.5 III	  \\
HR 1411     &   3.849 &  0.584 &  0.394 &  0.393  &   K0 IIIb	  \\
HR 1457     &   0.860 &  0.955 &  0.814 &  0.373  &   K5 III	  \\
HR 1577     &   2.690 &  0.937 &  0.775 &  0.307  &   K3 II	  \\
HR 2985     &   3.570 &  0.573 &  0.379 &  0.398  &   G8 IIIa	  \\
HR 3003     &   4.848 &  0.895 &  0.735 &  0.451  &   K5 III	  \\
HR 3249     &   3.518 &  0.914 &  0.758 &  0.371  &   K4 III	  \\
HR 3800     &   4.552 &  0.561 &  0.349 &  0.375  &   G8.5 III	  \\
% HR 3815     &   5.410 &  0.473 &  0.304 &  0.372  &   G8 IIIv	  \\
HR 4057     &   1.980 &  0.689 &  0.457 &  0.373  &   K1 IIIb	  \\
HR 4166     &   4.720 &  0.512 &  0.297 &  0.477  &   G2 IIa	  \\
HR 4392     &   4.989 &  0.610 &  0.416 &  0.396  &   G7.5 IIIa:  \\
HR 4695     &   4.969 &  0.717 &  0.485 &  0.516  &   K0 IIIb	  \\
HR 4883     &   4.932 &  0.437 &  0.186 &  0.416  &   G0 IIIp	  \\
HR 5681     &   3.490 &  0.587 &  0.346 &  0.410  &   G8 III	  \\
HR 5854     &   2.640 &  0.715 &  0.572 &  0.445  &   K2 IIIb	  \\
HR 5947     &   4.150 &  0.751 &  0.570 &  0.414  &   K2 IIIab	  \\
HR 5997     &   4.316 &  0.522 &  0.285 &  0.448  &   G3 II-III   \\
HR 6603     &   2.760 &  0.719 &  0.553 &  0.451  &   K2 III	  \\
HR 7328     &   3.760 &  0.579 &  0.390 &  0.430  &   G9 III	  \\
HR 7479     &   4.386 &  0.489 &  0.259 &  0.471  &   G1 III	  \\
HR 7525     &   2.711 &  0.936 &  0.762 &  0.292  &   K3 II	  \\
HR 7949     &   2.460 &  0.627 &  0.415 &  0.425  &   K0 III	  \\
HR 8551     &   4.790 &  0.640 &  0.420 &  0.418  &   K0 III	  \\
\enddata							 
\tablenotetext{a}{Taken from Perry et al.\ (1987).} 
\end{deluxetable}						 

\clearpage

% \end{document}

\startlongtable								 
\begin{deluxetable}{lcccc}					 
% \tabletypesize{\footnotesize} 				 
\tablecaption{\Gaia\/ Astrometry and Photometry, and Interstellar Reddening}
\tablewidth{0pt}						 
\tablehead{							 
\colhead{Star} & 						 
\colhead{Parallax} & 						 
\colhead{$G$} & 						 
\colhead{$BP-RP$} & 						 
\colhead{$E(B-V)$} \\						 
\colhead{ } & 							 
\colhead{[mas]} & 						 
\colhead{[mag]} & 						 
\colhead{[mag]} & 						 
\colhead{[mag]} 						 
}								 
\startdata							 
%               Plx     e       G       BP-RP   E(B-V)
\noalign{\medskip} 
\multispan5{\hfil Carbon-Deficient Red Giants \hfil} \\	
\noalign{\smallskip} 
CD $-$28 75  &  0.4986$\pm$0.0665 & 8.2987 & 1.8248 & 0.006       \\
HD 17232     &  0.8088$\pm$0.0638 & 8.0382 & 1.8924 & 0.025       \\ 
HR 885       &  6.5994$\pm$0.0865 & 5.1716 & 1.1194 & 0.043       \\ 
HD 18636     &  2.7035$\pm$0.0235 & 7.3743 & 1.1088 & 0.010       \\ 
HD 20090     &  1.7151$\pm$0.0422 & 7.6706 & 1.4126 & 0.026       \\ 
HR 1023      &  2.0972$\pm$0.0939 & 6.1664 & 1.0321 & 0.139       \\ 
BD +5 593    &  1.4583$\pm$0.0467 & 8.8350 & 1.5829 & 0.228       \\ 
HR 1299      &  3.5592$\pm$0.0230 & 6.1213 & 1.2286 & 0.007       \\ 
HD 28932     &  1.9047$\pm$0.0533 & 7.6307 & 1.2312 & 0.050       \\ 
HD 30297     &  2.1367$\pm$0.0745 & 8.2256 & 1.3422 & 0.139       \\ 
HD 31274     &  2.6616$\pm$0.0263 & 6.8408 & 1.1674 & 0.008       \\ 
HD 31869     &  1.0438$\pm$0.0251 & 9.0202 & 1.1305 & 0.011       \\ 
HD 36552     &  2.2443$\pm$0.0249 & 7.8348 & 1.0587 & 0.011       \\ 
HD 40402     &  1.5522$\pm$0.0341 & 8.3367 & 1.1283 & 0.014       \\ 
HD 49960     &  1.6239$\pm$0.0275 & 8.0318 & 1.2412 & 0.044       \\ 
HD 54627     &  1.6871$\pm$0.0290 & 8.4834 & 1.1887 & 0.047       \\ 
HD 56438     &  2.3080$\pm$0.0289 & 7.7341 & 1.3098 & 0.055       \\ 
HD 67728     &  1.2132$\pm$0.0301 & 7.1902 & 1.2959 & 0.048       \\ 
HD 78146     &  1.7226$\pm$0.0368 & 8.2095 & 1.3580 & 0.089       \\ 
HD 82595     &  1.8694$\pm$0.0380 & 7.8857 & 1.2194 & 0.047       \\ 
HD 91622     &  1.8444$\pm$0.0702 & 7.8740 & 1.3384 & 0.026       \\ 
HR 4154      &  3.9740$\pm$0.0299 & 5.8126 & 1.1531 & 0.023       \\ 
HD 94956     &  1.6564$\pm$0.0508 & 8.1743 & 1.1617 & 0.046       \\ 
HD 102851    &  1.3248$\pm$0.0447 & 8.4399 & 1.2874 & 0.060       \\ 
CD $-$37 7613&  1.0594$\pm$0.0397 & 9.5697 & 1.1968 & 0.058       \\ 
HD 105783    &  1.3947$\pm$0.0473 & 8.7378 & 1.4393 & 0.073       \\ 
37 Com       &  4.6981$\pm$0.2586 & 4.4196 & 1.3136 & 0.010       \\
HD 120170    &  1.4789$\pm$0.0701 & 8.7479 & 1.1751 & 0.030       \\ 
HR 5188      &  3.7775$\pm$0.0759 & 5.3562 & 1.6874 & 0.131       \\ 
HD 132776    &  1.7687$\pm$0.0444 & 8.4605 & 1.3656 & 0.076       \\ 
HD 146116    &  2.0063$\pm$0.0492 & 7.3473 & 1.2972 & 0.107       \\ 
HR 6476      &  3.6937$\pm$0.1082 & 5.3506 & 1.3883 & 0.125       \\ 
HR 6757      &  4.3580$\pm$0.0566 & 5.9747 & 1.2850 & 0.130       \\ 
HR 6766      &  9.8252$\pm$0.3403 & 4.2189 & 1.1515 & 0.007       \\ 
HR 6791      &  9.2013$\pm$0.1416 & 4.7065 & 1.0946 & 0.019       \\ 
HD 188028    &  2.2604$\pm$0.0458 & 7.4084 & 1.2602 & 0.100       \\ 
HD 198718    &  1.7095$\pm$0.0440 & 8.3491 & 1.1477 & 0.021       \\ 
HD 201557    &  1.2113$\pm$0.0422 & 8.9142 & 1.2243 & 0.019       \\ 
HD 204046    &  1.1602$\pm$0.0722 & 8.6612 & 1.3101 & 0.049       \\ 
HD 207774    &  1.3876$\pm$0.1160 & 8.6543 & 1.1484 & 0.041       \\ 
HD 215974    &  1.7723$\pm$0.0511 & 8.1979 & 1.1992 & 0.054       \\ 
\noalign{\smallskip} 
\multispan5{\hfil Barium Red Giants \hfil} 		        \\ 
\noalign{\smallskip} 
%                  Plx   e       G       BP-RP   E(B-V)         \\ 
HD 4084       &  2.5712$\pm$0.0554 & 8.2927 & 1.2823 & 0.010      \\ 
HD 4395       & 11.1382$\pm$0.0470 & 7.4924 & 0.8954 & 0.002      \\ 
HR 774        &  6.7929$\pm$0.0709 & 5.4525 & 1.3260 & 0.011      \\ 
HD 19014      &  1.4191$\pm$0.0232 & 7.5465 & 1.8653 & 0.084      \\ 
HD 20394      &  2.4187$\pm$0.0910 & 8.4484 & 1.2412 & 0.092      \\ 
HD 24035      &  4.6116$\pm$0.1006 & 8.2024 & 1.2731 & 0.042      \\ 
HD 26886      &  2.7112$\pm$0.0652 & 7.7281 & 1.1250 & 0.044      \\ 
HD 29370      &  1.5816$\pm$0.0279 & 9.0462 & 1.1815 & 0.010      \\ 
HD 31487      &  3.0355$\pm$0.1312 & 7.6813 & 1.4909 & 0.095      \\ 
HD 32712      &  2.6213$\pm$0.0257 & 8.2169 & 1.2753 & 0.012      \\ 
HD 36598      &  3.2140$\pm$0.0352 & 7.7137 & 1.3177 & 0.028      \\ 
HD 43389      &  2.0493$\pm$0.0461 & 7.8749 & 1.5432 & 0.069      \\
HR 2392       &  7.8396$\pm$0.1463 & 5.9974 & 1.1517 & 0.003      \\
HD 49641      &  2.0232$\pm$0.0401 & 6.7837 & 1.3833 & 0.028      \\
HD 50082      &  4.1443$\pm$0.0418 & 7.1856 & 1.1401 & 0.011      \\
HD 62017      &  1.6433$\pm$0.0610 & 8.5875 & 1.0426 & 0.030      \\
HD 65854      &  5.4912$\pm$0.0492 & 8.1641 & 1.1079 & 0.012      \\
HD 82221      &  2.6114$\pm$0.0333 & 7.5541 & 1.4015 & 0.034      \\
HD 84678      &  1.6498$\pm$0.0291 & 8.5698 & 1.5245 & 0.204      \\
HD 88035      &  1.4561$\pm$0.0501 & 8.8829 & 1.1885 & 0.031      \\
HD 88562      &  1.5268$\pm$0.0784 & 8.1199 & 1.5252 & 0.041      \\
HD 89175      &  3.8752$\pm$0.0348 & 7.4491 & 1.1846 & 0.044      \\
HD 91208      &  3.9627$\pm$0.0679 & 7.8020 & 1.0853 & 0.028      \\
HD 92626      &  3.4440$\pm$0.0379 & 6.7579 & 1.3305 & 0.027      \\
HR 4474       &  7.2676$\pm$0.0992 & 5.8697 & 1.1408 & 0.011      \\
HD 107541     &  4.1893$\pm$0.0507 & 9.1422 & 1.1479 & 0.024      \\
HR 5058       & 14.5400$\pm$0.2753 & 4.7947 & 1.2178 & 0.005      \\
HD 120620     &  3.6027$\pm$0.0930 & 9.3574 & 1.1797 & 0.023      \\
HD 123949     &  2.3231$\pm$0.0763 & 8.3298 & 1.4387 & 0.049      \\
HD 178717     &  2.6300$\pm$0.0490 & 6.4707 & 1.9294 & 0.162      \\
HD 183915     &  2.7905$\pm$0.0400 & 6.9457 & 1.3954 & 0.084      \\
HD 199435     &  3.5910$\pm$0.0277 & 7.9785 & 1.2628 & 0.067      \\
HD 199939     &  2.6616$\pm$0.0360 & 7.1338 & 1.2741 & 0.037      \\
HD 201657     &  3.1769$\pm$0.1145 & 7.7117 & 1.2917 & 0.068      \\
HD 201824     &  2.1605$\pm$0.0512 & 8.6855 & 1.1689 & 0.031      \\
$\zeta$ Cap   &  7.3533$\pm$0.4817 & 3.4260 & 1.1502 & 0.005      \\
HD 204886     &  2.2651$\pm$0.0375 & 7.8308 & 1.3246 & 0.016      \\
HD 205011     &  6.7342$\pm$0.0486 & 6.1636 & 1.1566 & 0.018      \\
HD 211594     &  3.5820$\pm$0.0704 & 7.8269 & 1.2030 & 0.048      \\
HD 219116     &  1.5836$\pm$0.0439 & 9.0319 & 1.1403 & 0.017      \\
CPD $-$64 4333&  2.1027$\pm$0.0736 & 9.3313 & 1.2374 & 0.013      \\
\noalign{\smallskip} 
\multispan5{\hfil Field Red Giants \hfil} \\
\noalign{\smallskip} 
%                   Plx     e       G       BP-RP    E(B-V) 	        \\
% 10-oct-19: Note: added the E(B-V)'s at request of referee
HR 373        &  13.4527$\pm$0.1345  & 5.2376 &  1.1523 & 0.003  \\
HR 1030       &  17.1066$\pm$0.3775  & 3.2567 &  1.1403 & 0.001  \\
HR 1327       &   9.4908$\pm$0.0960  & 5.0182 &  0.9874 & 0.004  \\
HR 1346       &  22.6234$\pm$0.4614  & 3.2900 &  1.2221 & 0.001  \\
HR 1373       &  19.0632$\pm$0.3699  & 3.4155 &  1.1604 & 0.001  \\
HR 1409       &  20.3130$\pm$0.4261  & 3.1587 &  1.2315 & 0.001  \\
HR 1411       &  21.4183$\pm$0.3457  & 3.4931 &  1.1451 & 0.001  \\
HR 2985       &  23.6199$\pm$0.3954  & 3.2379 &  1.1903 & 0.001  \\
HR 3003       &   9.1500$\pm$0.2963  & 4.2185 &  1.7024 & 0.003  \\
HR 3249       &  11.0443$\pm$0.6561  & 2.9266 &  1.6155 & 0.003  \\
HR 3800       &  18.1458$\pm$0.2345  & 4.2327 &  1.0733 & 0.002  \\
HR 4166       &   5.2136$\pm$0.4108  & 4.4153 &  0.9701 & 0.015  \\
HR 4392       &   5.8742$\pm$0.1937  & 4.6768 &  1.1159 & 0.010  \\
HR 4695       &   9.7577$\pm$0.2536  & 4.5190 &  1.3541 & 0.004  \\
HR 4883       &  11.4933$\pm$0.1828  & 4.6729 &  0.8637 & 0.003  \\
HR 5681       &  26.7797$\pm$0.3806  & 3.0531 &  1.2199 & 0.003  \\
HR 5854       &  39.3696$\pm$0.8514  & 2.1248 &  1.3238 & 0.001  \\
HR 5947       &  14.2898$\pm$0.2149  & 3.6599 &  1.4006 & 0.010  \\
HR 5997       &  10.8678$\pm$0.3421  & 4.0135 &  1.0270 & 0.030  \\
HR 6603       &  40.0945$\pm$0.6752  & 2.2781 &  1.9369 & 0.001  \\
HR 7328       &  27.0371$\pm$0.1921  & 3.4397 &  1.1574 & 0.003  \\
HR 7479       &   8.5307$\pm$0.1848  & 4.1175 &  0.9445 & 0.012  \\
HR 7949       &  43.1769$\pm$0.9384  & 2.0227 &  3.1413 & 0.001  \\
HR 8551       &  21.0039$\pm$0.2419  & 4.4191 &  1.2298 & 0.001  \\
\enddata
\end{deluxetable}

\end{document}